\documentstyle[12pt,fleqn,math,texdraw]{article}
\begin{document}
\parskip=4mm
\parindent=2.5em
\renewcommand{\theequation}{\arabic{section}.\arabic{equation}}
\renewcommand{\thefootnote}{*}
\thispagestyle{empty}
\noindent March 1994 \hfill{UNIL-TP-1/94}

\vspace*{1cm}
\begin{center}
{\LARGE\bf Fermion emission in a two-dimensional black hole
space-time\footnotemark}

\vspace{8mm}

{\large G. Wanders}\\[4mm]
Institut de physique th\'eorique, Universit\'e de Lausanne\\
CH-1015 Lausanne, Switzerland\\
{\small (e-mail: BANANTHA@ULYS.UNIL.CH)}

\footnotetext{Work supported by the Swiss National Science Foundation}

\vspace{12mm}

{\small{\bf Abstract}}
\end{center}

\noindent{\small We investigate massless fermion production by a
two-dimensional dilatonic
black hole. Our analysis is based on the Bogoliubov transformation relating
the outgoing fermion field observed outside the black hole horizon to the
incoming field present before the black hole creation. It takes full account
of the fact that the transformation is neither invertible nor unitarily
implementable. The particle content of the outgoing radiation is specified by
means of inclusive probabilities for the detection of sets of outgoing
fermions and antifermions in given states. For states localized near the
horizon these probabilities characterize a thermal equilibrium state. The way
the probabilities become thermal as one approaches the horizon is discussed in
detail.}

\vfill{\noindent PACS numbers: 04.62.+v, 04.70.+s, 11.10.Kk.

$\mbox{}$}
\newpage
\setcounter{page}{1}
\section{Introduction}

\hspace*{2.5em}In recent years the study of two-dimensional black holes has
been a field of
intense activity. In particular, the work of Callan~{\it et al.}~\cite{Cal} on
a black hole formed by infalling matter in the linear dilaton vacuum of
two-dimensional dilaton gravity has triggered a number of investigations on the
evaporation of a black hole due to the emission of Hawking radiation and its
back reaction on the geometry~\cite{Int2}. Whereas the black hole of
Ref.~\cite{Cal} provides
also the stage of the present work, we are not concerned with the back
reaction problem. Our purpose is to perform a detailed analysis of the
asymptotics of a massless fermion field living on a black hole space-time in
the case of a simple metric. In contrast with the more realistic
four-dimensional problem, the fermion field can be constructed explicitly and
the properties of the radiation generated by the black hole in the incoming
fermion vacuum are obtained in a simple way. The thermal aspects of this
radiation are readily identified.

The case of a massless scalar field has already been discussed in~\cite{Int3}.
We prefer to play with fermions for the following reasons. While the equations
of motion of massless fields are easily solved in two dimensions, the infrared
pathologies of such fields are worse in two than in higher dimensions. They
are especially acute for a scalar field: even the free field on flat space
cannot be defined properly without resorting to a Hilbert space with
indefinite metric~\cite{Int4}. These intricacies have been ignored
in~\cite{Int3} and we want to avoid them. The situation is easier with a
fermion field. The only difficulty one encounters is that the Bogoliubov
transformation relating the outgoing to the incoming fields is not unitarily
implementable. This is due to infrared and ultraviolet singularities and means
that an infinite number of fermion-antifermion pairs are created in the
incoming vacuum. Consequently there is no vector in the Fock space of the
outgoing particles representing the incoming vacuum. The implementability of
the in-out Bogoliubov transformation has been explicitly or tacitly assumed in
many discussions of the Hawking radiation~\cite{Int5}; it does not hold for
fermions in two dimensions.

One of our goals is to show that the outgoing particle content of the incoming
vacuum can be analyzed in spite of this non-implementability. This has already
been established in a previous work~\cite{Int6} for the case of a curvature
with compact support. Our strategy is to use inclusive probabilities for the
detection of collections of outgoing particles in given one-particle states.
Due to the exclusion principle these probabilities are easily computed for
fermions. In the boson case, only mean numbers of outgoing particles in given
states could be obtained easily; it would be harder to find how these numbers
are distributed.

Being compelled to work only with inclusive probabilities, there is no
temptation to talk about particles which end up in the black hole singularity.
An advantage of our approach is that it deals exclusively with observable
outgoing particles detected outside the black hole horizon.

This article is organized as follows. In Section~2 our space-time is equiped
with convenient incoming and ougoing coordinate frames, the Dirac equation is
solved in these frames and the Bogoliubov transformation relating the out- and
incoming fields is constructed. The inclusive probabilities specifying the
outgoing particle content of the incoming vacuum are defined in Section~3. We
show in Section~4 that the outgoing radiation observed at late times, near the
black hole horizon, is a thermal Hawking radiation. Finally, the relatively
slow rate at which the radiation becomes thermal as one approaches the horizon
is estimated in Section~5. Our conclusions are summarized in
Section~6 and technicalities are collected in four Appendices.

\setcounter{equation}{0}
\section{Fermion field in a black hole background}

\hspace*{2.5em}We want to study the creation of massless fermions and
antifermions in a two-dimensional space-time with metric
\begin{eqnarray}
{\rm d}s^2&=&{\rm e}^{\Omega(z)}\,{\rm d}z^+\,{\rm d}z^-,\nonumber\\[4mm]
\Omega(z) &=&\left\{\begin{array}{ll}-\log(-\lambda^2z^+z^-)
&z^+<z_0^+,\\[4mm]
\d -\log\left({M\over \lambda}-\lambda^2z^+(z^-+\Delta)\right) \qquad &
z^+>z_0^+.
\end{array}\right.
\end{eqnarray}
$z^\pm$ are Kruskal light-cone coordinates, $z^\pm=z^0\pm z^1$; the space we
are considering corresponds to the wedge $z^+>0,\;z^-<0$. It is the space-time
of a black hole formed in a dilaton vacuum by a pulse of infalling
massless matter propagating along the line $z^+=z_0^+$~\cite{Cal} (Fig.~1). $M$
is the mass of the black hole and $\lambda$ is a cosmological constant fixing a
length scale. A horizon is formed at $z^-=-\Delta=-(M/\lambda^3z_0^+)$,
$z^+>z_0^+$: the space-time is flat for $z^+<z_0^+$ and far away from the
horizon, $z^+<z_0^+$, $z^-\to -\infty$.

We follow the same strategy as in~\cite{Int6} and describe our fermion field in
terms
of suitably defined incoming and outgoing coordinates $x$ and $y$. The incoming
variables are given by
\begin{equation}
x^+={1\over \lambda}\ln{z^+\over z_0^+},\qquad x^-=-{1\over \lambda}
\ln\left(-{z^-\over \Delta}\right).
\end{equation}
The sector $(z^+>0,\;z^-<0)$ is mapped onto $\R^2$, the infalling pulse
follows the line $x^+=0$ and the horizon is at $x^-=0$, $x^+>0$. The
metric~(2.1) becomes
\begin{eqnarray}
{\rm d}s^2&=&{\rm e}^{\Omega_{\rm in}(x)}\,{\rm d}x^+\,{\rm d}x^-,\nonumber\\
\Omega_{\rm in}(x) &=&\left\{\begin{array}{ll}0 & x^+<0,\\
 -\ln\left(1-{\rm e}^{\lambda x^-}(1-{\rm e}^{-\lambda x^+})\right) \qquad
&x^+>0.
\end{array}\right.
\end{eqnarray}

The incoming variables are minkowskian before the infalling matter creates the
black hole.

The outgoing coordinates $(y^+,y^-)$ become minkowskian asymptotically, for
large values of $x^+$:
\begin{equation}
y^+=x^+,\qquad y^-=-{1\over \lambda}\,\ln({\rm e}^{-\lambda x^-}-1),\qquad
x^-<0.
\end{equation}
They map the exterior of the horizon, $x^-<0$, on the whole $(y^+,y^-)$ plane,
the image of the horizon itself being at $y^-=\infty$. The metric~(2.1) takes
the form
\begin{eqnarray}
{\rm d}s^2&=&{\rm e}^{\Omega_{\rm out}(y)}\,{\rm d}y^+\,{\rm d}y^-,
\nonumber\\[4mm]
\Omega_{\rm out}(y) &=&\left\{\begin{array}{ll}-\ln(1+{\rm e}^{\lambda y^-})
&y^+<0,\\[4mm]
 -\ln\left(1+{\rm e}^{-2\lambda y^1})\right) \qquad &y^+>0.
\end{array}\right.
\end{eqnarray}
It becomes static after the black hole has been created. The scalar curvature
is given by
\begin{equation}
R(y)=\left\{\begin{array}{ll}0&y^+<0,\\[4mm]
 \d {\lambda^2\over (1+{\rm e}^{2\lambda y^1})}, \qquad &y^+>0.
\end{array}\right.
\end{equation}
It is equal to $\lambda^2$ on the horizon ($y^-=+\infty$, $y^+>0$) and
decreases to zero at fixed positive $y^+$ for $y^-\to-\infty$. Our choice of
the outgoing coordinates is such that on the axis $y^1=0$ the curvature is
reduced to half its value on the horizon. We notice that the cosmological
constant $\lambda$ is the only parameter appearing in the forms~(2.3) and (2.5)
of the line element. The black hole mass $M$ has been evacuated by our choice
of origins. Our in- and outgoing coordinates are essentially the same as the
variables $y^\pm$ and $\sigma^\pm$ used in~\cite{Int3}.

We are interested in the right-moving fermions and antifermions produced
outside the horizon in the incoming fermionic vacuum as a result of the
formation of the black hole. This requires the knowledge of the right-moving
component of the fermion field. Let $\psi(x)$ designate this component in the
incoming frame and with respect to a zweibein parallel to the coordinate axis.
As shown in~\cite{Int6}, the Dirac equation for massless fermions in arbitrary
coordinates leads to:
\begin{equation}
\partial_+\left(\exp\left({1\over 4}\Omega(x)\right)\,\psi(x)\right)=0.
\end{equation}
This implies that our $\psi(x)$ has the following form:
\begin{equation}
\psi(x)=\exp\left(-{1\over 4}\Omega_{\rm in}(x)\right)\,\psi_{\rm in}(x^-)
\end{equation}
where $\psi_{\rm in}(x^-)$ is a free, flat-space, right-moving massless
fermion field. It coincides with $\psi(x)$ for $x^+<0$.

Similarly, the same right-moving component is given in the outgoing frame by
\begin{equation}
\hat\psi(y)=\exp\left(-{1\over 4}\Omega_{\rm out}(y)\right)\,\psi_{\rm
out}(y^-)
\end{equation}
where $\psi_{\rm out}$ is again a free field which coincides with $\hat{\psi}$
for finite values of $y^-$ in the limit of large light-cone time $y^+$.

The Fourier components of $\psi_{\rm in}(x^-)$ and $\psi_{\rm out}(y^-)$ are
the annihilation and creation operators of the in- and outgoing particles.
Explicitly,
\begin{eqnarray}
\psi_{\rm in}(x^-) &=&{1\over \sqrt{2\pi}}\int_0^\infty{\rm d}k\left[a_{\rm
in}(k)\,{\rm e}^{-{\rm i}kx^-}+b_{\rm in}^\dagger(k)\,{\rm e}^{{\rm i}kx^-
}\right]\nonumber\\
&&\\
\psi_{\rm out}(y^-) &=&{1\over \sqrt{2\pi}}\int_0^\infty{\rm d}k\left[a_{\rm
out}(k)\,{\rm e}^{-{\rm i}ky^-}+b_{\rm out}^\dagger(k)\,{\rm e}^{{\rm i}ky^-}
\right]\nonumber
\end{eqnarray}
The outgoing particles are localized outside the horizon whereas the
localization of the incoming particles is unrestricted.

The fields $\psi(x)$ and $\hat{\psi}(y)$ are related by the transformation law
of the components of a spinor field under a change of coordinate and zweibein
frame. According to~\cite{Int6} this law gives
\begin{equation}
\hat{\psi}(y)=\left[{{\rm d}y^+(x^+)\over{\rm d}x^+}\Big/
{{\rm d}y^-(x^-)\over {\rm d}x^-}\right]^{1\over 4}\psi(x).
\end{equation}
This relation
combined with (2.10) defines the Bogoliubov transformation connecting the
in- and outgoing creators and annihilators
\begin{eqnarray}
a_{\rm out}(k) &=&\int_0^\infty{\rm d}k\left[K_1(k,k')a_{\rm in}(k')+
K_2(k,k')b_{\rm out}^\dagger(k')\right]\nonumber\\[-2mm]
&&\\[-2mm]
b_{\rm out}^\dagger(k) &=&\int_0^\infty{\rm d}k\left[K_3(k,k')a_{\rm in}(k')+
K_4(k,k')b_{\rm out}^\dagger(k')\right]\nonumber
\end{eqnarray}

As in~\cite{Int6} the kernels $K_i$ are obtained from a single function $U$
\begin{equation}\begin{array}{ll}
K_1(k,k')=U(k,k'),&K_2(k,k')=U(k,-k'),\\[4mm]
K_3(k,k')=U(-k,k'),\qquad &K_4(k,k')=U(-k,-k'),
\end{array}\end{equation}
with
\begin{equation}
U(k,k')={1\over 2\pi}\int_{-\infty}^0{\rm d}x^-
({\rm d}y^-(x^-)/{\rm d}x^-)^{1\over 2}\,\exp\left[{\rm i}(ky^-(x^-)-k'x^-
)\right].
\end{equation}
Quite similar expressions hold for the Bogoliubov coefficients specifying the
in-out correspondence of a massless scalar field~\cite{Int3}. As in that case,
the
integral in (2.14) is a $\beta$ function:
\begin{equation}
U(k,k')={1\over 2\pi\lambda}\,B\left({{\rm i}\over \lambda}(k-
k')+\epsilon,{1\over 2}-{\rm  i}{k\over \lambda}\right).
\end{equation}

Two points must be emphasized about the transformation (2.12):
\begin{enumerate}
\item[(i)] Due to the fact that $\psi_{\rm out}(y^-)$ describes particles
which are observed outside the horizon, it is determined by the restriction of
$\psi_{\rm in}(x^-)$ to $x^-<0$. This is the reason why the integral in (2.14)
extends over negative values of $x^-$ only. As a consequence, the
transformation (2.12) is not invertible: the matrix $K$ formed by the $K_i$ is
such that $KK^\dagger=\id$ whereas $K^\dagger K\neq\id$. To obtain an
invertible transformation, auxiliary modes, localized in the interior region of
the black hole, must be introduced~\cite{Int3}.
\item[(ii)] Even if (2.12) is extended to an invertible mapping, this
mapping is not unitarily implementable. This means in particular that there is
no vector $\Omega_{\rm in}$, in the Fock space ${\cal F}_{\rm out}$ of the
outgoing particles, which represents the incoming vacuum. Therefore the
outgoing particle content of the incoming vacuum cannot be obtained from the
analysis of an $\Omega_{\rm in}$ in ${\cal F}_{\rm out}$.
\end{enumerate}

The non-implementability of the Bogoliubov transformation is due to the fact
that the kernels $K_2$ and $K_3$ have infinite Hilbert-Schmidt norms. To see
this, one observes that (2.13) and (2.14) imply
\begin{eqnarray}
\left|K_2(k,k')\right|^2 &=&{1\over 2\pi}\,{1\over k+k'}
\,{\cosh(\pi k'/\lambda)\over \cosh(\pi k/\lambda)}
\,{1\over \sinh(\pi(k+k')/\lambda)}\nonumber\\[-2mm]
&&\\[-2mm]
&>&{1\over 2\pi}\,{1\over k+k'}\,\exp(-\pi k/\lambda).\nonumber
\end{eqnarray}
The last inequality shows that the integral
\begin{equation}
\Vert K_2\Vert_{\rm HS}^2 =\int_0^\infty{\rm d}k\int_0^\infty{\rm
d}k'\,\left|K_2(k,k')\right|^2
\end{equation}
has an infrared divergence in $k$ and $k'$ and an ultraviolet divergence in
$k'$. The infrared divergence comes from the fact that we are dealing with
massless fermions whereas the ultraviolet divergence is produced by the sharp
cutoff at the horizon $x^-=0$ of the integral in (2.14). The physical meaning
of the divergence of (2.17) is that an infinite number of fermion-antifermion
pairs is created by the black hole. We notice that a similar ultraviolet
divergence appears in the Hilbert-Schmidt norm of the off-diagonal Bogoliubov
coefficients of a massless scalar field in a 4-dimensional black hole space-
time~\cite{Int5}.

We show in the following Sections that the physics of massless fermion
production can be analysed in spite of the fact that the Bogoliubov
transformation is neither invertible nor implementable. In particular, one can
do without unobservable modes located beyond the horizon.

We close this Section by displaying the two quantities which are most easily
obtained from the Bogoliubov coefficients: the mean values of the
right-flowing current and energy-momentum in the incoming vacuum. As
in~\cite{Int6}, $K_3(k,k')=K_2^*(k,k')$ and this implies again that the mean
current vanishes identically: fermions and antifermions are
created in pairs, on top of each other. The expectation value of the outgoing
component $\theta_{--}^{\rm out}$ of the energy momentum tensor is given by
\begin{equation}
\theta_{--}^{\rm out}={{\rm i}\over 2}\;:\psi_{\rm out}^\dagger
\stackrel{\d \leftrightarrow}{\partial_{y^-}}\psi_{\rm out}:_{\rm out},
\end{equation}
where $\;:\quad:_{\rm out}\;$ means normal ordering with respect to the
outgoing
creators and annihilators. One finds
\begin{equation}
\bra\Omega_{\rm in},\theta_{--}^{\rm out}(y^-)\Omega_{\rm in}\ket
={\lambda^2\over 48\pi}\left(1-{1\over \left(1+{\rm e}^{\lambda y^-
}\right)^2}\right),
\end{equation}
as in the case of bosons~\cite{Cal}. This expression agrees with the
asymptotic form of the energy momentum tensor obtained from the trace anomaly.

As one approaches the horizon ($y^-\to\infty$) this density tends
exponentially to the constant value $\lambda^2/48\pi$. One may ask if this is
the rate at which the radiation accompanying the black hole becomes a thermal
Hawking radiation. The next Sections show that the situation is not as simple
as that.

\setcounter{equation}{0}
\section{Outgoing particles in the incoming vacuum}

\hspace*{2.5em}As one cannot construct an incoming vacuum state vector in the
outgoing Fock space, it is impossible to define the exclusive probability of
finding a state formed by a given set of outgoing particles in the incoming
vacuum. A vector $\Omega_{\rm in}$ exists only in the incoming Fock space
${\cal F}_{\rm in}$ and it is in this space that one is forced to work. What is
well defined in this space is the inclusive probability of detecting a given
configuration of outgoing particles, irrespective of the presence or absence of
further outgoing particles. The inclusive probability $W[f]$ of one outgoing
fermion in a state $f$ has been examined in~\cite{Int6}:
\begin{equation}
W[f]=\bra\Omega_{\rm in},\,P[f]\Omega_{\rm in}\ket,
\end{equation}
$f\in L^2(\R_+,{\rm d}k),\quad\Vert f\Vert=1$. $P[f]$ is the projector on the
subspace of the incoming Fock space in which the outgoing one-particle state
$f$ is occupied. The advantage of dealing with fermions is that this projector
has a simple form:
\begin{equation}
P[f]=a_{\rm out}^\dagger[f^*]a_{\rm out}[f].
\end{equation}
In our notation the operators $a_{\rm out}[f]$ and $a_{\rm out}^\dagger[f]$ are
linear in $f$: $\d a_{\rm out}[f]=\int_0^\infty{\rm d}k\,a_{\rm out}(k)f(k)$, a
similar expression for $a_{\rm out}^\dagger[f]$ being obtained by replacing
$a_{\rm out}(k)$ by $a_{\rm out}^\dagger(k)$. The adjoint of $a_{\rm out}[f]$
is given by $\left(a_{\rm out}[f]\right)^\dagger=a_{\rm out}^\dagger[f^*]$.
In conjunction with this definition of the smeared creators and annihilators,
it is convenient to define the bilinear form
\begin{equation}
(f_i,f_j)=\int_0^\infty{\rm d}k\,f_i(k)f_j(k).
\end{equation}
The scalar product of $f_i$ and $f_j$ is then equal to $(f_i^*,f_j)$. Using
(3.2) and (2.12) one finds that the inclusive probability (3.1) is given by
\begin{equation}
W[f]=(f,K_2K_2^\dagger f^*).
\end{equation}

One may ask if there are observers who can really measure the probability
$W[f]$. This question is discussed in Appendix~A: the answer is positive.

As we want to know if the particles radiated by the black hole form a thermal
ensemble we need not only the one-particle inclusive probability~(3.1) but
also the inclusive probabilities of arbitrary collections of outgoing
particles. This means that we have to find an expression for
$W_{N,\bar{N}}[f_1,\dots,f_N;\bar{f}_1,\dots,\bar{f}_{\bar{N}}]$, the
inclusive probability of $N$ outgoing fermions and $\bar{N}$ outgoing
antifermions in the orthonormal states $\{f_1,\dots,f_N\}$,
$\{\bar{f}_1,\dots,\bar{f}_{\bar{N}}\}$,
$(f_i^*,f_j)=(\bar{f}_i^*,\bar{f}_j)=\delta_{ij}$. Besides the projector $P[f]$
defined in (3.2) we need the projector $\bar{P}[\bar{f}]$ associated with the
antifermion one-particle state $\bar{f}$:
\begin{equation}
\bar{P}[\bar{f}]=b_{\rm out}^\dagger[\bar{f}^*]b_{\rm out}[\bar{f}].
\end{equation}
Due to the orthonormality of the $f$ and $\bar{f}$, the $P[f_i]$ and
$\bar{P}[\bar{f}_j]$ commute among themselves and their products are
projectors. Consequently,
\begin{equation}
W_{N,\bar{N}}[f_1,\dots;\bar{f}_1,\dots]=\bra\Omega_{\rm
in},\,\prod_{i=1}^NP[f_i]\,\prod_{j=1}^{\bar{N}}\bar{P}[\bar{f}_j]\,\Omega_{\rm
in}\ket.
\end{equation}

The projectors $P$ and $\bar{P}$ can be expressed explicitly in terms of the
fermion fields, for instance:
\begin{equation}
P[f]=\int{\rm d}y\int{\rm d}y'\tilde{f}^*(y)\psi_{\rm out}^\dagger(y)\psi_{\rm
out}(y')\tilde{f}(y'),
\end{equation}
where $\tilde{f}$ is the wave function defined in (4.1). This shows that
$W_{N,\bar{N}}$ is obtained from an $(N+\bar{N})$-point function of the
outgoing fermion field in the incoming vacuum. We prefer to work with the
expressions (3.2) and (3.5) of the projectors. After insertion into (3.6) one
finds that $W_{N,\bar{N}}$ can be written as a sum of products of
$(N+\bar{N})$ contractions of pairs of outgoing creators and annihilators.
An explicit formula for $W_{N,\bar{N}}$ can be written down. One defines
$(N+\bar{N})$ operators $\alpha_i$ which either annihilate a fermion or
create an antifermion:
\begin{equation}
\hspace*{-1cm}\alpha_i=a_{\rm out}[f_i],\quad i=1,\dots,N;\qquad
\alpha_i=b_{\rm out}^\dagger[\bar{f}_{i-N}^*],\quad i=N+1,\dots,N+\bar{N}.
\end{equation}
Each permutation $\pi$ of $(1,\dots,N+\bar{N})$ generates a term of
$W_{N,\bar{N}}$ in which $\alpha_i^\dagger$ is paired with $\alpha_{\pi_i}$.
We show in Appendix~B that $W_{N,\bar{N}}$ can be written as a determinant:
\begin{equation}
W_{N,\bar{N}}[f_1,\dots;\bar{f}_1,\dots]=\sum_\pi(-1)^{\sigma(\pi)}
\prod_{i=1}^N\mbox{
\raisebox{-3mm}{\parbox[t]{1cm}{\btexdraw \drawdim mm \linewd 0.2
\rlvec (0 -2)\rlvec (5 0)\rlvec (0 2)\textref h:L v:B
\htext (-1 0){$\alpha_i^\dagger$}
\htext (4 0){$\alpha_{\pi_i}$} \move (10 5)\etexdraw}}}
\prod_{j=N+1}^{N+\bar{N}}(-1)\mbox{
\raisebox{-3mm}{\parbox[t]{1cm}{\btexdraw \drawdim mm \linewd 0.2
\rlvec (0 -2)\rlvec (6 0)\rlvec (0 2)\textref h:L v:B
\htext (-1 0){$\alpha_{\pi_j}$}
\htext (5 0){$\alpha_j^\dagger$} \move (10 5)\etexdraw}}}.
\end{equation}

To evaluate the contractions we have to express the
out-operators in (3.8) in terms of in-operators. Equations (2.12) give

\begin{eqnarray}
a_{\rm out}[f] &=&a_{\rm in}[fK_1]+b_{\rm in}^\dagger[fK_2]\nonumber\\[-2mm]
&&\\[-2mm]
b_{\rm out}^\dagger[f^*] &=&a_{\rm in}[f^*K_3]+b_{\rm in}^\dagger[f^*K_4].
\nonumber
\end{eqnarray}

Therefore, if $i\leq N$ and $j\leq N$,

\begin{eqnarray}
\mbox{
\raisebox{-3mm}{\parbox[t]{1cm}{\btexdraw \drawdim mm \linewd 0.2
\rlvec (0 -2)\rlvec (5 0)\rlvec (0 2)\textref h:L v:B
\htext (-1 0){$\alpha_i^\dagger$}
\htext (4 0){$\alpha_j$} \move (10 5)\etexdraw}}}
&=&\bra\Omega_{\rm in},\,b_{\rm in}[f_i^*K_2^*]\,b_{\rm
in}^\dagger[f_jK_2]\Omega_{\rm in}\ket\nonumber\\
&=&(f_j,\,K_2K_2^\dagger f_i^*)
\end{eqnarray}

For the other values of $i$ and $j$ one finds
\begin{eqnarray}
\mbox{
\raisebox{-3mm}{\parbox[t]{1cm}{\btexdraw \drawdim mm \linewd 0.2
\rlvec (0 -2)\rlvec (5 0)\rlvec (0 2)\textref h:L v:B
\htext (-1 0){$\alpha_i^\dagger$}
\htext (4 0){$\alpha_j$} \move (10 5)\etexdraw}}}&=& (f_i,K_1K_1^\dagger
f_j^*)\;=\;\delta_{ij}-(f_i,K_2K_2^\dagger f_j^*)\quad\mbox{if }\quad i>N,\quad
j>N,\nonumber\\
\mbox{
\raisebox{-3mm}{\parbox[t]{1cm}{\btexdraw \drawdim mm \linewd 0.2
\rlvec (0 -2)\rlvec (5 0)\rlvec (0 2)\textref h:L v:B
\htext (-1 0){$\alpha_i^\dagger$}
\htext (4 0){$\alpha_j$} \move (10 5)\etexdraw}}}&=&(f_i^*,K_3K_1^\dagger
f_j^*)\qquad\mbox{if }\quad i\leq N,\quad j> N,\\
\mbox{
\raisebox{-3mm}{\parbox[t]{1cm}{\btexdraw \drawdim mm \linewd 0.2
\rlvec (0 -2)\rlvec (5 0)\rlvec (0 2)\textref h:L v:B
\htext (-1 0){$\alpha_i^\dagger$}
\htext (4 0){$\alpha_j$} \move (10 5)\etexdraw}}}&=&(f_i,K_3K_1^\dagger f_j)
\qquad\mbox{if }\quad i> N,\quad j\leq N.\nonumber
\end{eqnarray}
In these formulae $f_i$ stands for $\bar{f}_{i-N}$ if $i>N$. The relations
$K_1K_1^\dagger +K_2K_2^\dagger=\id$, $K_3=K_2^*$, $K_4=K_1^*$ have been used
as well as the orthonormality of the $\bar{f}$.

The remaining contractions are obtained from (3.11) and (3.12) by
\begin{equation}
\mbox{
\raisebox{-3mm}{\parbox[t]{1cm}{\btexdraw \drawdim mm \linewd 0.2
\rlvec (0 -2)\rlvec (5 0)\rlvec (0 2)\textref h:L v:B
\htext (-1 0.5){$\alpha_i$}
\htext (4 -0.5){$\alpha_j^\dagger$} \move (10 5)\etexdraw}}}=\delta_{ij}-
\mbox{
\raisebox{-3mm}{\parbox[t]{1cm}{\btexdraw \drawdim mm \linewd 0.2
\rlvec (0 -2)\rlvec (5 0)\rlvec (0 2)\textref h:L v:B
\htext (-1 -0.5){$\alpha_j^\dagger$}
\htext (4 0.5){$\alpha_i$} \move (10 5)\etexdraw}}}
\end{equation}

An important result is that all
inclusive probabilities are assemblies of two building blocks:
$(f_i,K_2K_2^\dagger f_j^*)$ for the fermion-fermion and
antifermion-antifermion correlations, and $(f_i, K_1K_3^\dagger f_j)$ and its
complex conjugate for the fermion-antifermion correlations. These are examined
in detail in the following Sections. Notice that the term ``correlation" has a
special meaning here: it refers to correlations as they appear in a set of
inclusive probabilities.

The formula (3.9) implies a relation expressing the general fermion-antifermion
symmetry of the inclusive probabilities:
\begin{equation}
W_{N,\bar{N}}[f_1,\dots f_N;\bar{f}_1,\dots,\bar{f}_{\bar{N}}]=
W_{\bar{N},N}[\bar{f}_1,\dots,\bar{f}_{\bar{N}};f_1,\dots f_N].
\end{equation}

As an illustration, we give a list of the simplest inclusive probabilities:
\begin{eqnarray}
W_{1,0}[f] &=&W_{0,1}[f]\;=\;\left(f,Lf^*\right)\nonumber\\[4mm]
W_{2,0}[f_1,f_2] &=&W_{1,0}[f_1]W_{1,0}[f_2]-\left|\left(f_1,L
f_2^*\right)\right|^2\nonumber\\[4mm]
W_{0,2}[f_1,f_2] &=&W_{2,0}[f_1,f_2] \\[4mm]
W_{1,1}[f,\bar{f}] &=&W_{1,0}[f]W_{0,1}[\bar{f}]+\left|\left(f,K_1K_3^\dagger
\bar{f}\right)\right|^2\nonumber\\[4mm]
W_{3,0}[f_1,f_2,f_3] &=&W_{1,0}[f_1]W_{1,0}[f_2]W_{1,0}[f_3]
-W_{1,0}[f_1]\left|\left(f_2,L f_3^*\right)\right|^2\nonumber\\
&&-W_{1,0}[f_2]\left|\left(f_3,L f_1^*\right)\right|^2
-W_{1,0}[f_3]\left|\left(f_1,L f_2^*\right)\right|^2\nonumber\\
&&+\left\{\left(f_1,Lf_2^*\right)\left(f_2,Lf_3^*\right)
\left(f_3,Lf_1^*\right)+\mbox{ compl. conj.}\right\}\nonumber
\end{eqnarray}
The notation $L=K_2K_2^\dagger$ has been used. The antisymmetry of
$K_1K_3^\dagger$ has been taken into account in $W_{1,1}$: it results from
$K_1K_3^\dagger+K_2K_4^\dagger=0$ (consequence of $KK^\dagger=\id$) and
$K_2=K_3^*$, $K_1=K_4^*$. Notice that this antisymmetry
implies $W_{1,1}[f,f]=W_{1,0}[f]W_{0,1}[f]=\left(W_{1,0}[f]\right)^2$. The
inclusive probability of observing a fermion and an antifermion in the same
state is simply the square of the probability of detecting one particle in
this state. This fact is clearly related to the vanishing of the mean outgoing
current.

It is worth noting that all we have done in this Section is not restricted to
the case of a non-implementable transformation~(2.12). The definition~(3.6)
and the formula~(3.9) remain valid if the transformation is implementable. In
that case, we could also compute exclusive probabilities of outgoing states
containing a given finite set of particles. As an infinite number of particles
is created if the transformation is non-implementable, all exclusive
probabilities vanish in our case.

\setcounter{equation}{0}
\section{Thermal radiation near the horizon}

\hspace*{2.5em}Manageable expressions for $(f_i,K_2K_2^*f_j^*)$ and
$(f_i,K_1K_3^*f_j)$, the inclusive probabilities' building blocks, are obtained
if they are written in terms of the Fourier transforms of the momentum space
wave functions $f$:
\begin{equation}
\tilde{f}(y^-)={1\over \sqrt{2\pi}}\int_0^\infty{\rm d}k\,f(k)\,{\rm e}^{{\rm
i}ky^-}.
\end{equation}
This is a fortunate circumstance because it gives easy access to the dependence
of the inclusive probabilities on the localization of the states $f_i$. The
form of the building blocks we shall use has been derived in~\cite{Int6}. As it
will play a key role, its derivation is briefly repeated here. The definition
(2.13--14) of $K_2$ gives
\begin{eqnarray}
\left(K_2K_2^\dagger\right)(k_1,k_2)&=&{{\rm i}\over 4\pi^2}
\int_{-\infty}^0{\rm d}x_1\int_{-\infty}^0{\rm d}x_2
\,{\sqrt{y'(x_1)y'(x_2)}\over x_1-x_2+{\rm i}\epsilon}\nonumber\\[4mm]
&&\qquad\qquad\exp\left[{\rm i}\left(k_1y(x_1)-k_2y(x_2)\right)\right].
\end{eqnarray}
Here and in what follows, $x$ stands for $x^-$ and $y(x)=y^-(x)$ as defined in
(2.4); $y'(x)={\rm d}y/{\rm d}x$. Switching to $y$-variables, equation~(4.2)
gives
\begin{equation}
\hspace*{-1cm}\left(f_i,K_2K_2^\dagger f_j^*\right)={1\over 2\pi{\rm i}}
\int_{-\infty}^{+\infty}{\rm d}y_1\int_{-\infty}^{+\infty}{\rm d}y_2
\,\tilde{f}_j^*(y_1)\,{\sqrt{x'(y_1)x'(y_2)}\over x(y_1)-x(y_2)-
{\rm i}\epsilon}\,\tilde{f}_i(y_2),
\end{equation}
where $x(y)$ is the inverse of $y=y(x)$ and $x'(y)$ is its derivative.
This integral is difficult to handle because of the singularity at $y_1=y_2$.
In fact we can get rid of this singularity. It follows from the
definition~(4.1) that $f_i$ is regular in the upper half-plane. Consequently
the following integral vanishes:
\begin{equation}
{1\over 2\pi{\rm i}}\int{\rm d}y_1\int{\rm d}y_2
\,f_j^*(y_1)\,{1\over y_1-y_2-{\rm i}\epsilon}\,f_i(y_2)=0.
\end{equation}
Subtracting this integral from (4.3) we obtain
\begin{equation}
\left(f_i,K_2K_2^\dagger f_j^*\right)=\left(\tilde{f}_j^*,G
\tilde{f}_i\right),
\end{equation}
with
\begin{equation}
G(y_1,y_2)={1\over 2\pi{\rm i}}\left({\sqrt{x'(y_1)x'(y_2)}\over x(y_1)-x(y_2)-
{\rm i}\epsilon}-{1\over y_1-y_2-{\rm i}\epsilon}\right).
\end{equation}

As announced, this kernel is regular at $y_1=y_2$ and the ${\rm i}\epsilon$
terms can be dropped. A similar computation shows that
\begin{equation}
\left(f_i,K_1K_3^\dagger f_j\right)={1\over 2{\rm i}\pi}
\left(\tilde{f}_i,G \tilde{f}_j\right).
\end{equation}
Equations (4.5) and (4.6) tell us that all inclusive probabilities can be
expressed in terms of a single functional
\begin{equation}
A(g_1,g_2)=(g_1,Gg_2).
\end{equation}
The ingredients of $W_{n,\bar{n}}$ are obtained by identifying the functions
$g_i$ either with wave functions $\tilde{f}_i$ or their complex conjugates.

The formulae (4.5)--(4.8) do not depend on the specific form of $x(y)$, the
function relating the outgoing coordinate $y^-$ to the incoming $x^-$. In the
case of our black hole metric, this function is
\begin{equation}
x(y)=-{1\over \lambda}\,\ln \left(1+{\rm e}^{-\lambda y}\right)
\end{equation}
according to (2.4). It depends on the single parameter $\lambda$ and, as a
consequence of (4.6), the kernel $G$ obeys a simple scaling law under a change
of $\lambda$:
\begin{equation}
G_{\lambda'}(y_1,y_2)={\lambda'\over \lambda}\,G_\lambda
\left({\lambda'\over \lambda}y_1,{\lambda'\over \lambda}y_2\right).
\end{equation}
This implies that inclusive probabilities for two values of $\lambda$ are
related by
\begin{equation}
W_{N,\bar{N}}^{(\lambda')}\left(f_1,\dots;\bar{f}_1,\dots\right)=
W_{N,\bar{N}}^{(\lambda)}\left(f'_1,\dots;\bar{f}_1^{\,'},\dots\right).
\end{equation}
The new wave functions $f'_i$ are obtained from $f_i$ by scaling:
\begin{equation}
\tilde{f}_i^{\,'}(y)=\left({\lambda\over \lambda'}\right)^{1\over 2}\tilde{f}_i
\left({\lambda\over \lambda'}y\right).
\end{equation}
The same relation holds for $\tilde{\bar{f}_i}^{\,'}$ and $\tilde{\bar{f}_i}$.
If we consider states $f_i$ and $\tilde{\bar{f}_i}$ localized around positive
values $a_i$ and $\bar{a}_i$ of $y$, with widths $\delta_i$ and
$\bar{\delta}_i$, their inclusive probability for a large value $\lambda'$ of
the cosmological  constant is equal to the probability for a smaller value
$\lambda$ ($\lambda<\lambda'$) of broader states $\tilde{f}_i^{\,'}$ and
$\tilde{\bar{f}_i}^{\,'}$ (widths
$\delta'_i=(\lambda'/\lambda)\delta_i\dots$) localized closer to the horizon,
at larger values of $y$, $a'_i=(\lambda'/\lambda)a_i,\dots$.

We are now ready for the main topic of this work: the thermal nature of the
radiation accompanying our black hole. A signal of a thermal radiation emerges
from the preceding formulae if one applies them to states localized near the
horizon i.e.~states detected at late times $\tau$ by the observer described
in Appendix~A. The wave functions of such states are concentrated on large
positive values of $y$: $\lambda y\gg 1$ (remember that $y$ stands for the
outgoing light-cone coordinate $y^-$). For these large values of $y$, the
function $x(y)$ of equation~(4.9) can be approximated by
\begin{equation}
x(y)\simeq -(1/\lambda)\exp(-\lambda y).
\end{equation}
The replacement of $x(y)$ by the right-hand side of this equation
will be denoted as the horizon approximation. In this approximation the kernel
$G$ defined in (4.6) becomes translation invariant and is given by
\begin{equation}
G_{\rm H}(y_1-y_2)={1\over 2\pi{\rm i}}\left({\lambda\over
2\sinh\left({\lambda\over 2}(y_1-y_2)\right)}-{1\over y_1-y_2}\right).
\end{equation}
$G_{\rm H}$ is close to $G$ if $\lambda y_i\gg 1$, $i=1,2$. If the functions
$g_i$ in
(4.8) are localized at large
positive values of $y$, $G$ can be replaced by $G_{\rm H}$ in (4.8). This
leads to an approximate value $A_{\rm H}$ of $A$. We shall evaluate it now and
discuss the quality of the approximation in Section~5.

Before we start, a general and important observation has to be made. The
states $f_i$ and $\bar{f}_i$ being positive energy states, the wave functions
$\tilde{f}_i$ and $\tilde{\bar{f}_i}$ are analytic functions of $y$, regular
in the upper half-plane (cf.~(4.1)). Their complex conjugates are regular in
the lower half-plane. These analyticity properties play a prominent role in
the following discussion. As the arguments $g_1$ and $g_2$ of the functional
$A$ stand either for a wave function or its complex conjugate, these functions
are regular in the upper or lower half-plane. The knowledge of $A[g_1,g_2]$
when at least one of the functions $g_1$ and $g_2$ is regular in $\Im\,y>0$,
gives full control of all the contractions listed in (3.11-12). We shall always
assume that $g_2$ has this  property.

$G_{\rm H}(y_1-y_2)$ is a meromorphic function of $y_2$ with poles at
$y_2=y_1\pm{\rm i}(2n\pi/\lambda)$, $n=1,\dots$. The integral over $y_2$ in
$A_{\rm H}$ is thus equal to an infinite sum of residues of upper half-plane
poles:
\begin{equation}
A_{\rm H}(g_1,g_2)=\int_{-\infty}^{+\infty}{\rm d}y\,g_1(y)\,g_2^{(1)}(y),
\end{equation}
where $g_2^{(1)}$ is an auxiliary function:
\begin{equation}
g_2^{(1)}(y)=\sum_{n=1}^\infty(-1)^{n+1}g_2\left(y+{\rm
i}(2n\pi/\lambda)\right).
\end{equation}

If $g_2=\tilde{f}_i$ and $g_1$ is the complex conjugate of the wave function
$\tilde{f}_j$, it is readily seen that the integral in (4.15) is equal to
\begin{equation}
A_{\rm H}(\tilde{f}_j^*,\tilde{f}_i)=\int_0^\infty{\rm d}k\,
{f_j^*(k)f_i(k)\over 1+{\rm e}^{\beta k}},\qquad \beta=2\pi/\lambda.
\end{equation}
If, on the other hand, $g_2=\tilde{f}_i$ and $g_1=\tilde{f}_j$, the integral
in (4.15) vanishes, its integrand being regular in $\Im\,y>0$, and
\begin{equation}
A_{\rm H}(\tilde{f}_i,\tilde{f}_j)=0.
\end{equation}

These results have far-reaching implications. As $A(\tilde{f}_i,\tilde{f}_j)$
determines the fermion-anti\-fer\-mion correlations of the inclusive
probabilities, equation~(4.18) tells us that these correlations vanish in the
horizon approximation. As a consequence, the inclusive probabilities factorize
in this approximation:
\begin{equation}
W_{N,\bar{N}}^{\rm H}[f_1,\dots;\bar{f}_1,\dots]=
W_{N,0}^{\rm H}[f_1,\dots]\;W_{0,\bar{N}}^{\rm H}[\bar{f}_1,\dots].
\end{equation}

As $W_{0,N}^{\rm H}[f_1,\dots,f_N]=W_{N,0}^{\rm H}[f_1,\dots,f_N]$
(cf.~(3.14)), we can restrict our discussion of the horizon approximation to
the fermionic inclusive probabilities. These probabilities have the form
\begin{equation}
W_{N,0}^{\rm H}[f_1,\dots]=\sum_\pi(-1)^{\sigma(\pi)}\prod_{i=1}^NA_{\rm
H}\left(\tilde{f}_i^*,f_{\pi_i}\right).
\end{equation}

The simple form of $A_{\rm H}(f_j^*,f_i)$ in (4.17) is due to the fact that the
horizon approximation $G_{\rm H}$ of $G$ is translation invariant and
meromorphic with equally spaced complex poles. Exactly the same mechanism has
been encountered in~\cite{Int6} in the case of a point-like curvature. It
suggests that we are dealing with a thermal radiation of inverse temperature
$\beta$. To turn this observation into a firm statement we have to show that
all horizon inclusive probabilities $W^{\rm H}$ are inclusive probabilities
of a thermal radiation. We could do this by proving that the
probabilities~(4.20) characterize a KMS state~\cite{Int8}. Our strategy is
more conservative and we show that the probabilities $W^{\rm H}$ coincide with
the probabilities $W^{\rm th}$ produced by a thermal density matrix $\rho$.

As we want to deal with a well defined density matrix we consider a
system of free massless right-moving fermions and antifermions which is
localized on a finite interval $[y_1,y_2]$ with periodic boundary conditions,
$y_2=y_1+(2\pi/\epsilon)$, $\epsilon>0$. The one-particle energy eigenstates
are
$\phi_n(y)=\sqrt{\epsilon/2\pi}\,\exp(-{\rm i}n\epsilon y)$, $n=0,1,\dots$ and
an arbitrary one-particle state has the form
\begin{equation}
\psi(y)=\sum_{n=0}^\infty\gamma_n\,\phi_n(y),\qquad
\Vert\psi\Vert^2=\sum_n|\gamma_n|^2=1.
\end{equation}

The density matrix $\rho$ is defined on the Fock space ${\cal H}$ of our
fermion system:
\begin{equation}
\rho=Z^{-1}\,\exp\left(-\beta\sum_n n\epsilon \left(c_n^\dagger c_n+d_n^\dagger
d_n
\right)\right),
\end{equation}
where $c_n$ annihilates a fermion in the state $\phi_n$ and $d_n$ annihilates
an antifermion in the same state. This density matrix determines thermal
inclusive probabilities $W^{\rm th}$. The inclusive probability
$W_{N,\bar{N}}^{\rm th}[\psi_1,\dots,\bar{\psi}_1,\dots]$ of
detecting $N$ fermions in the orthonormal states $\psi_1,\dots,\psi_N$ and
$\bar{N}$ antifermions in the states $\bar{\psi}_1,\dots,\bar{\psi}_{\bar{N}}$
is given by
\begin{equation}
W_{N,\bar{N}}^{\rm th}[\psi_1,\dots,\bar{\psi}_1,\dots]=
\Tr\left(\rho\prod_{i=1}^NP_{\rm th}[\psi_i]
\prod_{j=1}^{\bar{N}}\bar{P}_{\rm th}[\bar{\psi}_j]\right).
\end{equation}
$P_{\rm th}[\psi]$ is the projector onto the subspace of ${\cal H}$ in which
the one-fermion state $\psi$ is occupied:
\begin{equation}
P_{\rm th}[\psi]=\sum_{n,n'}\gamma_n^*\gamma_{n'}c_n^\dagger c_{n'}.
\end{equation}
A similar definition holds for the antifermion projector $\bar{P}_{\rm th}$.

It is readily seen that the thermal probabilities $W^{\rm th}$ obey the same
factorization (4.19) as the horizon probabilities $W^{\rm H}$. Working out the
right-hand side of (4.23) one discovers that the thermal probabilities are
given by precisely the same formula (4.20) as the horizon probabilities if
$A_{\rm H}$ is replaced by a thermal functional $A_{\rm th}$:
\begin{equation}
A_{\rm th}(\psi_j^*,\psi_i)=\d \sum_n{\gamma_n^{j*}\gamma_n^i\over
1+{\rm e}^{\beta n\epsilon}}.
\end{equation}
The proof of these results is outlined in Appendix~C. $A_{\rm th}$ is obviously
a discretized version of $A_{\rm H}$ as given in (4.17). As the
normalization constant $Z$ does not appear in the final expressions of the
thermal inclusive probabilities $W^{\rm th}$, we may take the infinite volume
limit $y_1\to-\infty$, $y_2\to+\infty$, $\epsilon\to 0$. In this limit the
thermal and horizon probabilities coincide. This means that the probabilities
$W^{\rm H}$ really describe a thermal radiation as it is observed by means of
inclusive detections of given arbitrary finite configurations of quanta.

These results do not imply that the black hole radiation is entirely thermal.
The exact probabilities $W$ coincide with $W^{\rm H}$ only for states close to
the horizon. Therefore it is only the radiation localized near the horizon
which is thermal.

\setcounter{equation}{0}
\section{Approach to the thermal radiation}

\hspace*{2.5em}We established in the preceding Section that the thermal nature
of the black
hole radiation is encoded in the horizon approximation of our inclusive
probabilities. This approximation is certainly a good one for states
sufficiently close to the horizon, that is for states detected sufficiently
late by the observer of Appendix~A. In this Section we want to obtain a
precise formulation of this statement. We ask how fast an
inclusive probability $W$ approaches its horizon approximation when its states
are shifted towards the horizon. To be specific, we take a collection of
states $f_i$ and $\bar{f}_i$ centred on finite values $a_i$ and $\bar{a}_j$ of
$y$ and we shift them to the right by a distance $y_0$:
$\tilde{f}_1(y)\to\tilde{f}_{y_0,1}(y)=\tilde{f}_1(y-y_0),\dots$. At fixed
$f_i$ and $\bar{f}_j$, the inclusive probability becomes a function of $y_0$:
\begin{equation}
W_{y_0}\left[f_1,\dots;\bar{f}_1,\dots\right]=
W\left[f_{y_0,1},\dots;\bar{f}_{y_0,1},\dots\right].
\end{equation}
The horizon approximation $G_{\rm H}$ of the kernel $G$ being translation
invariant
(cf.~(4.14)), the horizon approximation of $W_{y_0}$ is independent of $y_0$.
We expect that
\begin{equation}
\lim_{y_0\to\infty}W_{y_0}\left[f_1,\dots;\bar{f}_1,\dots\right]
=W^{\rm H}\left[f_1,\dots;\bar{f}_1,\dots\right]
\end{equation}
and we would like to know the decay law of the difference $W_{y_0}-W^{\rm H}$
at
large values of $y_0$. According to our previous results, this law is
determined by the behavior of the difference $\Delta A_{y_0}=A_{y_0}-A_{\rm H}$
where $A_{y_0}(g_1,g_2)=A(g_{y_0,1},g_{y_0,2})$. In view of the scaling law
(4.10) we can restrict our discussion to the case $\lambda=1$. We shall do
this; $\lambda$ is set equal to $1$ from now on. $\Delta A_{y_0}$ is given by
\begin{equation}
\Delta A_{y_0}(g_1,g_2) =\int_{-\infty}^{+\infty}{\rm d}y_1
\int_{-\infty}^{+\infty}{\rm d}y_2\,g_1(y_1-y_0)\Delta G(y_1,y_2)g_2(y_2-y_0),
\end{equation}
the explicit form of $\Delta G$ being
\begin{eqnarray}
\Delta G(y_1,y_2)&=&{1\over 2{\rm i}\pi}\left\{\left[(1+{\rm e}^{y_1})
(1+{\rm e}^{y_2})\right]^{-{1\over 2}}\left[\ln{1+{\rm e}^{-y_2}\over
1+{\rm e}^{-y_1}}\right]^{-1} \right.\nonumber\\
&&\qquad -2\left.\left[\sinh \left({1\over 2}(y_1-y_2)
\right)\right]^{-1}\right\}.
\end{eqnarray}

The test functions $g_1$ and $g_2$ in (5.3) are concentrated on large positive
values of $y_1$ and $y_2$, of the order of $y_0$. Inspection of (5.4) shows
that $\Delta G(y_1,y_2)$ is small for such values, proportional to
$\exp(-2y_0)$. This implies that the contribution to $\Delta A_{y_0}$ of the
central parts of $g_1$ and $g_2$ goes to zero exponentially if $y_0\to\infty$.
One may be tempted to conclude that $\Delta A_{y_0}$ itself decays
exponentially. This fails to be true because of a combination of two facts. In
the first place, $\Delta G$ behaves differently for large positive values and
large negative values of its arguments. The compensation between the first and
second terms in the right-hand side of (5.4), which is responsible for the
exponential decrease of $\Delta G(y_1,y_2)$ for large positive values of $y_1$
and $y_2$, is no longer at work for large negative values. The first term
varies as $(y_1-y_2)^{-1}$ and dominates the second term outside the diagonal
$y_1=y_2$. Secondly, the slow decrease of $\Delta G$ for $y_i\to-\infty$,
$i=1,2$, cannot
be overcome by a rapid decrease of the functions $g_i(y_i-y_0)$. These have
unavoidable tails, extending over the whole negative axis, which can decrease
fast, for instance like a large negative power, but not exponentially fast.
This is due to the regularity of each of
these functions, either in the upper or in the lower half-plane, this
regularity itself being a consequence of the fact that we are dealing with
positive energy states. The outcome of these circumstances is that the
asymptotics of $\Delta A_{y_0}$ as $y_0\to\infty$ is determined by the tails of
$g_1$ and $g_2$ on the negative $y$-axis; its decay is relatively slow and
depends crucially on the shape of these tails. There is no universal decay law
for $\Delta A_{y_0}$. The fact that the tails of wave functions have to be
taken into account has been noticed in~\cite{Int7}

After this qualitative discussion we derive an exact upper bound for $\Delta
A_{y_0}$ that is in agreement with our observations. We indicate the strategy
of the derivation here, the technicalities being relegated to the Appendices~C
and D. The upper bound is obtained in three steps. First, the expression
(5.3-4) is transformed and rewritten in terms of two new auxiliary functions
$g_i^{(2)}$, $i=1,2$. Then we derive an upper bound for $\Delta A_{y_0}$ in
terms of these functions. Finally, the bound is reformulated in terms of the
original functions $g_1$ and $g_2$.

The transformation of the expression (5.3--4) of $\Delta A_{y_0}$ is based on
the analyticity properties of the integrand in (5.3). The kernel $\Delta
G(y_1,y_2)$ is, at fixed real $y_1$, an analytic and periodic function of $y_2$
with branch points at $y_2=(2n+1){\rm i}\pi$, $n\in\Z$ (see Appendix~C). If, as
in Section~4, $g_2$ is regular in the upper half-plane, the integral over $y_2$
is equal to a sum of integrals along the cuts attached to the branch points
located in this half-plane. The integral over $y_1$ in the resulting expression
$\Delta A_{y_0}$ is submitted to a similar procedure and one finds
\begin{equation}
\Delta A_{y_0}[g_1,g_2]=\pm\int_{-\infty}^0{\rm d}y_1
\int_{-\infty}^0{\rm d}y_2\,\Delta G_2(y_1,y_2)g_1^{(2)}(y_1-y_0)
g_2^{(2)}(y_2-y_0)
\end{equation}
with
\begin{eqnarray}
\hspace*{-1cm}\Delta G_2[y_1,y_2] &=&{1\over {\rm i}\pi}\left[(1-{\rm e}^{y_1})
(1-{\rm e}^{y_2})\right]^{-{1\over 2}}\left\{P{1\over
\ln[({\rm e}^{-y_2}-1)/({\rm e}^{-y_1}-1)]}\right.\nonumber\\
&& \left.+{1\over 2}\left[{1\over
\ln[({\rm e}^{-y_2}-1)/({\rm e}^{-y_1}-1)]+{\rm i}\pi}
+\mbox{ compl. conj.}\right]\right\}
\end{eqnarray}

The auxiliary functions $g^{(2)}$ are defined by
\begin{equation}
g^{(2)}(y)=\sum_{n=0}^\infty(-1)^ng(y\pm(2n+1){\rm i}\pi).
\end{equation}
The $+$ sign has to be used if $g$ is regular in the upper half-plane whereas
the $-$ sign holds for a $g$ regular in the lower half-plane. Eq.~(5.5) holds
with a $+$ sign if $g_1$ and $g_2$ are regular in the same half-plane, and with
a $-$ sign if these functions are regular in opposite half-planes.

The main difference between equations (5.3) and (5.5) is that the double
integral is now restricted to negative values of $y_1$ and $y_2$. This shows
clearly that $\Delta A_{y_0}$ is determined by the left tails of
$g_i^{(2)}(y-y_0)$, $i=1,2$, which become smaller and smaller when $y_0$ tends
to infinity. The price one has to pay for this reduction of the domain of
integration is the principal part singularity of the new kernel $\Delta G_2$
at $y_1=y_2$.

Our bound for $\Delta A_{y_0}$ is based on formula~(5.5). As already mentioned,
the behavior of $\Delta A_{y_0}$ depends on the shape of the functions $g_1$
and $g_2$. Therefore we have to specify the class of functions we admit. If we
wish to obtain information on the behavior of the inclusive probabilities for
arbitrary states, the only restrictions we can impose on the wave functions
$\tilde{f}_i$ and $\tilde{\bar{f}_j}$ are that they should be
square-integrable, normalized and regular in the upper half of the $y$-plane.
For arbitrary states fulfilling these conditions the computations sketched in
Appendix~D lead to the final upper bound
\begin{equation}
\left|\Delta A_{y_0}(g_1,g_2)\right|<C\,L_{y_0}(g_1)+
D\,M_{y_0}(g_1)M_{y_0}(g_2),
\end{equation}
where $C$ and $D$ are constants and
\begin{equation}
L_{y_0}(g) = \inf_{Y>0}\left\{\kappa{\rm e}^{-Y}+\left[\int_{-\infty}^{Y-y_0}
{\rm d}y\left|g(y)\right|^2\right]^{1\over 2}\right\},\qquad \kappa=2/\pi^2.
\end{equation}
The same expression holds for $M_{y_0}(g)$ with $\kappa=1/\pi$.

As an illustration one can consider functions which have the form
\begin{equation}
g_i(y)={C\delta_i^{n-{1\over 2}}\over (y-a_i\pm{\rm i}\delta_i)^n}\,{\rm
e}^{{\rm i}\omega_iy},\qquad
n>{1\over 2}.
\end{equation}
Evaluating the infimum in (5.9) in the large $y_0$ limit one finds that
$L_{y_0}$ and $M_{y_0}$ behave as $\d y_0^{-(n-{1\over 2})}$. According to
(5.8) $\Delta A_{y_0}$ is bounded by an inverse power of $y_0$: an
exponential decay is excluded. This implies that the counting rates measured
by an observer at late times $\tau$ tend to the thermal counting rates,
as inverse powers of $\tau$, whereas the observed mean energy-momentum flux
(2.19) goes exponentially fast to its thermal limit. This contrast between
the behavior of a mean density and data
concerning individual particles is due to the long tails of positive energy
wave functions. Such tails are also present in the case of massive particles.

The bound~(5.8) is based on crude extimates and one may ask if an improved
bound could reveal a stronger decay of $\Delta A_{y_0}$. An approximate direct
evaluation of $\Delta A_{y_0}$ for functions~(5.10) with $n=1$ shows that,
although the bound (5.8) is substantially larger than $|\Delta A_{y_0}|$, it
reproduces correctly its behavior for large values of $y_0$.

\setcounter{equation}{0}
\section{Conclusions}

\hspace*{2.5em}We have obtained a complete description of the massless fermion
radiation
accompanying the CGHS black hole~\cite{Cal}, neglecting its back reaction on
the black hole. This description is based on
inclusive probabilities: it is exact and takes full account of the fact that
the automorphism relating the outgoing fields to the incoming ones is neither
invertible nor unitarily implementable. The non-implementability is due to an
infrared and ultraviolet singularity. The latter will survive if the fields
become massive. We deal only with observable outgoing particles, detected
outside the black hole horizon. Particles hitting the black hole singularity
do not appear in our discussion.

We have established the thermal nature of the radiation observed at late
times, near the horizon, by comparison with the inclusive probabilities of a
thermal equilibrium state. The approach to the asymptotic counting rates is
slow if compared to the behavior of the mean energy-momentum flux. This is due
to the long tails of positive energy wave functions which prevent a sharp
localization of the particles forming the outgoing radiation.

I thank F. Vendrell who checked a part of the computations presented in this
article.

\renewcommand{\thesection}{Appendix \Alph{section}}
\renewcommand{\theequation}{\Alph{section}.\arabic{equation}}
\setcounter{section}{0}
\setcounter{equation}{0}
\section{}
\vspace{-4mm}{\Large\bf Defining an observer}

We consider an observer with proper time $\tau$ whose worldline is a time-like
geodesic parametrized by $y=y(\tau)$ in the outgoing coordinate frame. The
observer carries a detector corresponding to the observable $Q^\dagger Q$ with
\begin{equation}
Q=\int{\rm d}\tau\,h(\tau)\hat{\psi}(y(\tau)),
\end{equation}
where $h$ is a square integrable test function. A similar detector is used
in~\cite{Int9} for a boson field.

The metric being static for $y^+>0$ (cf.~(2.5)), the observer's geodesic
becomes a straight line after he has met the infalling pulse creating the black
hole, say at time $\tau=0$. We choose an observer who is at rest for
$\tau>0$~(Fig.~1):
\begin{equation}
y^1(\tau)=Y, \qquad
y^0(\tau)=\exp\left(-{1\over 2}\Omega_{\rm out}(Y)\right)\tau\qquad
\mbox{if }\tau>0.
\end{equation}

If $Y$ is large and positive, $\d \Omega_{\rm out}(Y)$ is small
($\simeq\exp(-\lambda Y/2)$) and, according to (2.9), $\hat{\psi}(y(\tau))$ can
be
approximated by $\psi_{\rm out}(y^-(\tau))$. In fact, $\Omega_{\rm
out}(y(\tau))$ is exponentially small along the whole world line: it is nearly
a straight line also for $\tau<0$ and the approximation
$\hat{\psi}\simeq\psi_{\rm out}$ holds for all $\tau$. Therefore, in the limit
of large positive $Y$:
\begin{equation}
Q=\int_{-\infty}^{+\infty}{\rm d}\tau h(\tau)\psi_{\rm out}(\tau-Y).
\end{equation}
$Q$ becomes equal to the annihilator $a_{\rm out}[f]$ if $h$ is regular in the
upper half-plane and $\tilde{f}(y^-)=h(y^-+Y)$. In that case $Q^\dagger
Q=P[f]$ and our observer measures the inclusive probability $W[f]$. This
observer has to stay all the time in a region where space-time is flat or
nearly flat. If $h$ is concentrated on very large values of $\tau$, much
larger than $Y$, the observer explores the radiation at very large values of
$y^-$, close to the horizon located at $y^-=\infty$.

Notice that our choice excludes a test function $h$ with compact support: the
integral in (A.3) really extends over the whole axis. We may say that it is
impossible to have a strictly localized detector which counts particles without
creating some. However, such a detector can be reasonably well localized as
one may choose a function $h$ centered on some $\tau_0$ and decreasing as an
arbitrary large negative power of $(\tau-\tau_0)$.

\pagebreak
\setcounter{equation}{0}
\section{}
\vspace{-4mm}{\Large\bf Computing inclusive probabilities}

First we explain how one arrives at formula~(3.9). In terms of the $\alpha$
defined in (3.8), $W_{N,\bar{N}}$ has the form
\begin{equation}
W_{N,\bar{N}}=\left(
\Omega_{\rm in},\alpha_1^\dagger\alpha_1\dots\alpha_N^\dagger\alpha_N
\alpha_{N+1}\alpha_{N+1}^\dagger\dots\alpha_{N+\bar{N}}
\alpha_{N+\bar{N}}^\dagger\Omega_{\rm in}\right).
\end{equation}
The right-hand side is equal to a sum over all possible pairings
\raisebox{-3mm}{\parbox[t]{1cm}{\btexdraw \drawdim mm \linewd 0.2
\rlvec (0 -2)\rlvec (5 0)\rlvec (0 2)\textref h:L v:B
\htext (-1 0){$\alpha_i^\dagger$}
\htext (4 0){$\alpha_j$} \move (10 5)\etexdraw}} and
\raisebox{-3mm}{\parbox[t]{1cm}{\btexdraw \drawdim mm \linewd 0.2
\rlvec (0 -2)\rlvec (5 0)\rlvec (0 2)\textref h:L v:B
\htext (-1 0.5){$\alpha_i$}
\htext (4 -0.5){$\alpha_j^\dagger$} \move (10 5)\etexdraw}}, the order of the
factors being the same as in (B.1). Every $\alpha$ and $\alpha^\dagger$ is a
combination $(a_{\rm in}+b_{\rm in}^\dagger)$ and in each contraction the
operator on the right creates an incoming particle and the operator on the
left annihilates the same particle, therefore
\begin{equation}
\mbox{
\raisebox{-3mm}{\parbox[t]{1cm}{\btexdraw \drawdim mm \linewd 0.2
\rlvec (0 -2)\rlvec (5 0)\rlvec (0 2)\textref h:L v:B
\htext (-1 0){$\alpha_i^\dagger$}
\htext (4 0){$\alpha_j$} \move (10 5)\etexdraw}}}
=\left(\Omega_{\rm in},\alpha_i^\dagger\alpha_j\Omega_{\rm in}\right).
\end{equation}
The anticommutation relations of the $\alpha$ imply equation (3.13),
so all we have to know are the values of the contractions
\raisebox{-3mm}{\parbox[t]{1cm}{\btexdraw \drawdim mm \linewd 0.2
\rlvec (0 -2)\rlvec (5 0)\rlvec (0 2)\textref h:L v:B
\htext (-1 0){$\alpha_i^\dagger$}
\htext (4 0){$\alpha_j$} \move (10 5)\etexdraw}}. These are listed in
equations (3.11-12).

The general term of $W_{N,\bar{N}}$ being identified, it is easy to find the
sign in front of each of them. The result is displayed in equation~(3.9).

We turn now to the inclusive thermal probabilities defined in (4.22). As
already mention in Section~4, it is sufficient to consider the fermionic
probabilities $W_{N,0}^{\rm th}$. Using the orthonormality of the $\psi$,
these can be written as follows:
\begin{equation}
W_{N,0}^{\rm th}=(-1)^P\;\Tr\left(\rho c^\dagger[\psi_1^*\,]\dots
c^\dagger[\psi_N^*\,]c[\psi_1]\dots c[\psi_N]\right),
\end{equation}
where $P=N(N-1)/2$.

After insertion of the expansion (4.20) we get
\begin{equation}
W_{N,0}^{\rm th}=(-1)^P\sum_{\begin{array}{l}\scriptscriptstyle
n_1,n_2,\dots\\[-3mm]
\scriptscriptstyle
m_1,m_2,\dots\end{array}}\prod_{i=1}^N\gamma_{n_i}^{i^*}\gamma_{m_i}^i
\Tr\left(\rho c_{n_1}^\dagger\dots c_{n_N}^\dagger c_{m_1}\dots
c_{m_N}\right).
\end{equation}
The trace in the right-hand side is non-zero only if all the $n$ and $m$ are
distinct. This trace is evaluated in the Fock basis of the energy eigenstates.
Let $|\Phi\ket$ be one of these eigenstates: it is specified by the
collection $\{\phi_{\nu_1},\dots,\phi_{\nu_M}\}$ of one-fermion states which
are occupied in that state. The contribution of $|\Phi\ket$ to the trace is
non-zero only if $M=N$ and if the set $(\nu_1,\dots,\nu_N)$ is a permutation
of $(m_1,\dots,m_N)$. furthermore, as $|\Phi\ket$ is an eigenstate of $\rho$,
the set $(n_1,\dots, n_N)$ must also be a permutation of $(m_1,\dots,m_N)$. If
all these conditions are fulfilled one gets
\begin{equation}
\bra \Phi|\rho c_{n_1}^\dagger\dots c_{m_2}\dots |\Phi\ket = {1\over
Z}\exp\left(-\beta\epsilon\sum_{i=1}^Nn_i\right)\sum_\pi(-
1)^{\sigma(\pi)}\prod_j\delta_{m_j,n_{\pi_j}}
\end{equation}
where $\pi$ is the permutation of $(1,\dots,N)$ which transforms
$(n_1,\dots,n_N)$ into $(m_1,\dots,m_N)$. Using the value of the normalization
factor,
\begin{equation}
Z=\prod_{n=0}^\infty\left(1+{\rm e}^{-\beta n\epsilon}\right),
\end{equation}
we arrive at
\begin{equation}
W_{N,0}^{\rm th}=\sum_{n_1,\dots,n_N}\sum_\pi(-1)^{\sigma(\pi)}
\prod_{i=1}^N
{\gamma_{n_i}^{i^*}\gamma_{n_i}^{\pi_i}\over 1+{\rm e}^{\beta n_i\epsilon}}.
\end{equation}

To obtain the final result stated in Section~4, we have to commute the sum
over the $n$ and the product of the $i$. This is not really allowed because
the $n$ are not independent: the sum includes only non-coinciding $n$.
However, if the states $\psi$ have broadly distributed expansion coefficients
$\gamma_n$, the adjunction of sets $(n_1,\dots,n_N)$ with coinciding elements
does not modify the sum in a significant way and the $n$ can be treated as if
they were independent. This leads to an expression of the form (4.20) for
$W_{N,0}^{\rm th}$ with $A_{\rm H}$ replaced by $A_{\rm th}$, as given in
(4.17).

\setcounter{equation}{0}
\section{}
\vspace{-4mm}{\Large\bf Deriving formula (5.5) for the approach to the thermal
radiation}

At fixed real $y_1$, $\Delta G(y_1,y_2)$ defined in (5.3) has an analytic
continuation in the complex plane. The first term in (5.3) has poles at
$y_2=y_1+2n{\rm i}\pi$, $n\in\Z$, and cuts at $y_2=(2n+1){\rm i}\pi$,
$n\in\Z$. The second term is a meromorphic function: it has the same poles as
the first term with exactly opposite residues. Therefore $\Delta G$ has only
branch points. If these branch points are provided with cuts parallel to the
negative real $y_2$-axis, $\Delta G$ is periodic in the cut plane:
\begin{equation}
\Delta G(y_1,y_2+2{\rm i}\pi)=\Delta G(y_1,y_2).
\end{equation}
This periodicity in imaginary light-cone time might suggest that the radiation
is thermal everywhere. This is not the case because a thermal radiation
requires kernels which are both periodic and meromorphic: our $\Delta G$ is
periodic but has branch points.

If $g_2$ is regular in the upper half-plane, the contour of the $y_2$-integral
can be displaced into this half-plane. One obtains a sum of integrals along
the cuts of $\Delta G$. Due to the periodicity~(C.1) this sum can be written
as
\begin{equation}
\Delta A[g_1,g_2]=\int_{-\infty}^{+\infty}{\rm d}y_1
\int_{-\infty}^0{\rm d}y_2\,\Delta G_1(y_1,y_2)g_1(y_1)g_2^{(2)}(y_2)
\end{equation}
where
\begin{eqnarray}
\hspace*{-1cm}\Delta G_1[y_1,y_2] &=&{1\over 2{\rm i}\pi}\left[(1+{\rm e}^{-
y_1})
(1-{\rm e}^{-y_2})\right]^{-{1\over 2}}\left\{
{1\over \ln[({\rm e}^{-y_2}-1)/({\rm e}^{y_1}-1)]+{\rm i}\pi}\right.\nonumber\\
&&+\mbox{ compl. conj.}\Biggr\}
\end{eqnarray}
and $g_2^{(2)}(y)$ is given in (5.7). The $y_0$ appearing in (5.3) has been
dropped temporarily. The functions $g_1$ and $g_2$ do not play symmetric roles
in (C.2). A transformation of the $y_1$-integral restores this symmetry. For
real $y_2$ the new kernel $\Delta G_1(y_1,y_2)$ has poles at
$y_1=y_2+(2n+1){\rm i}\pi$, $n\in\Z$ and branch points at
$y_1=(2n+1){\rm i}\pi$, $n\in\Z$. If the cuts attached to these
branch points are drawn parallel to the negative $y_1$-axis, the kernel
$\Delta G_1$ is periodic in $y_1$:
\begin{equation}
\Delta G_1(y_1+2{\rm i}\pi,y_2)=\Delta G_1(y_1,y_2).
\end{equation}
Now it is the contour of the $y_1$-integral that is deformed into the
half-plane in which $g_1$ is regular. Notice that, $y_2$ being negative, the
poles of $\Delta G_1$ are on its cuts. One obtains therefore a sum of integrals
along these cuts, each of them exhibiting a principal part singularity. The
final result is displayed in formula~(5.5).

\setcounter{equation}{0}
\section{}
\vspace{-4mm}{\Large\bf Bounds for the approach to the thermal radiation}

In a first step we derive an upper bound for $\Delta A$ in terms of the
auxiliary functions $g_i^{(2)}$, $i=1,2$. This bound is based on the
square-integrability and analyticity properties of these functions. It follows
directly from the definition~(5.7) that $g^{(2)}$ is regular in the half-plane
$\Im\,y>-\pi$ ($\Im\,y<\pi$) if $g$ is regular in the upper (lower)
half-plane. Furthermore, $g^{(2)}$ is square integrable and its $L_2$ norm is
smaller than that of $g$. This is a consequence of the Fourier transformed
version of (5.7). For $g$ regular in the upper or lower half-plane,
\begin{equation}
\tilde{g}^{(2)}(k) =\theta(\pm k)\tilde{g}(k){\rm e}^{\mp\pi k}\left(1+{\rm
e}^{\mp2\pi k}\right)^{-1}.
\end{equation}
The asymptotic behavior and the singularities of $\Delta G_2$ are such that
the square integrability of $g_i^{(2)}$ alone does not ensure the convergence
of the double integral in~(5.5). We show how to circumvent this difficulty in
the case of the first term of $\Delta G_2$ in its definition~(5.6). We call
this term $H(y_1,y_2)$ and rewrite it as follows:
\begin{equation}
H(y_1,y_2)=P{1\over y_1-y_2}+\left[H(y_1,y_2)-P{1\over y_1-y_2}\right].
\end{equation}
The virtue of this decomposition is that the simple distribution
$P(y_1-y_2)^{-1}$ has precisely the same singularity as $H(y_1,y_2)$ at
$y_1=y_2$ and the same asymptotic behavior for large negative $y$. The square
bracket is therefore regular at $y_1=y_2$ and has a better asymptotic behavior
than $H$.

To obtain an upper bound for the contribution $\Delta A_1$ of the first term
in (D.2) to $\Delta A$, we use the upper half-plane analyticity of $g_2^{(2)}$.
It implies that
\begin{equation}
g_2^{(2)}(y_1)={1\over {\rm i}\pi}P\int_{-\infty}^{+\infty}{\rm d}y_2{1\over
y_1-y_2}g_2^{(2)}(y_2).
\end{equation}
Dropping $y_0$ again, this relation allows us to write
\begin{eqnarray}
\hspace*{-1cm}
\pm\Delta A_1&=&P\int_{-\infty}^0{\rm d}y_1\int_{-\infty}^0{\rm d}y_2
\,g_1(y_1){1\over y_1-y_2}\,g_2(y_2)\\
&=&{\rm i}\pi\int_{-\infty}^0{\rm d}y\,g_1^{(2)}(y)g_2^{(2)}(y)-
\int_{-\infty}^0{\rm d}y_1\int_0^\infty{\rm d}y_2\,{1\over y_1-y_2}\,
g_1^{(2)}(y_1)g_2^{(2)}(y_2).\nonumber
\end{eqnarray}

As an illustration of the technique leading to our bounds, we show how a bound
is obtained for the second term in~(D.4). Introducing polar coordinates in the
$(-y_1,y_2)$ plane, the modulus of this term becomes
\begin{equation}
\left|\int_0^{\pi\over 2}{\rm d}\theta{1\over \sqrt{2}\cos\left(\theta-
{\pi\over 4}\right)}\int_0^\infty{\rm d}\rho\,g_1^{(2)}(-\rho\cos\theta)
g_2^{(2)}(\rho\cos\theta)\right|.
\end{equation}
The Schwartz inequality gives the upper bound
\[\hspace*{-1cm}\int_0^{\pi\over 2}{\rm d}\theta{1\over
\sqrt{2}\cos\left(\theta-
{\pi\over 4}\right)}\left[\int_0^\infty{\rm d}\rho\,\left|
g_1^{(2)}(-\rho\cos\theta)\right|^2\right.\left.\int_0^\infty{\rm
d}\rho\,\left|
g_2^{(2)}(\rho\cos\theta)\right|^2\right]^{1\over 2}\]
\begin{equation}
\vspace{-4mm}
\qquad=C\,\Vert g_1^{(2)}\Vert_-\,\Vert g_2^{(2)}\Vert_+,
\end{equation}
where
\begin{equation}
\Vert g\Vert_\pm^2=\int_{-\infty}^{+\infty}{\rm d}y\,\theta(\pm y)\,|g(y)|^2,
\end{equation}
and the constant $C$ is given by
\begin{equation}
C=\int_0^{\pi\over 2}{\rm
d}\theta\,{1\over\sqrt{\sin\,2\theta}\,\cos\Bigl(\theta-{\pi\over 4}\Bigr)}.
\end{equation}

Equations (D.4) and (D.6) lead to the bound
\begin{equation}
|\Delta A_1|<\Vert g_1^{(2)}\Vert_-\left(\pi\Vert g_2^{(2)}\Vert_-+C\Vert
g_2^{(2)}\Vert_+\right).
\end{equation}

Other bounds of the same type could be derived. For instance, we could have
operated with $g_1^{(2)}$ rather than $g_2^{(2)}$ in (D.3). What has been lost
in (D.9) is the antisymmetry of $\Delta A_1$ under the exchange of
$g_1^{(2)}$ and $g_2^{(2)}$. We shall not try to improve the bound (D.9): all
we want is a simple bound for $|\Delta A_{y_0}|$ allowing a crude but
exact control of its decay as $y_0\to\infty$.

We shall not go through the tedious derivation of the bounds for the remaining
terms of $\Delta A$. We just mention one intricacy due to the square root
singularities of $\Delta G_2$ at $y_1=0$ and $y_2=0$ (cf.~(5.6)). Near the
origin we have to deal with integrands of the form
$\d(-y)^{-{1\over 2}}g^{(2)}(y)$. The Schwarz inequality can be applied in the
following way:
\begin{equation}
\left|\int^0{\rm d}y(-y)^{-{1\over 2}}g_i^{(2)}(y)\right|
<\left[\int^0{\rm d}y(-y)^{-{1\over 2}}\int^0{\rm d}y(-y)^{-{1\over 2}}
\left|g_i^{(2)}(y)\right|^2\right]^{1\over 2}.
\end{equation}
The second integral on the right-hand side converges because the analyticity
of $g_i^{(2)}$ implies that the square of its modulus is regular at the
origin. In the explicit computations, $\d (-y)^{-{1\over 2}}$ must be replaced
by a function which has the same behavior at $y=0$ but is square integrable
at $-\infty$. We shall use $\d (\exp(-y)-1)^{-{1\over 2}}$.

The final upper bound of $\Delta A$ in terms of the auxiliary functions
$g_i^{(2)}$ has the following form:
\begin{equation}
|\Delta A(g_1,g_2)|<CI(g_1)+DJ(g_1)J(g_2)
\end{equation}
where $C$ and $D$ are constants and
\begin{equation}\begin{array}{lcl}
I(g)&=&\left(\int_{-\infty}^0{\rm d}y\left|g^{(2)}(y)\right|^2\right)^{1\over
2}\\
J(g)&=&\left(\int_{-\infty}^0{\rm d}y\left({\rm e}^{-y}-1\right)^{-{1\over
2}}\left|g^{(2)}(y)\right|^2\right)^{1\over 2}
\end{array}
\end{equation}

Our final task is to estimate these norms in terms of the original functions
$g$. It is convenient to rewrite the sum in the definition (5.7) of $g^{(2)}$
as an integral:
\begin{equation}
g^{(2)}(y)={1\over 4\pi}\int_{-\infty}^{+\infty}{\rm d}y'\,{1\over
\cosh\left({1\over 2}(y-y')\right)}\,g(y').
\end{equation}
This expression leads to
\begin{equation}
I^2(g)=\int_{-\infty}^{+\infty}{\rm d}y\int_{-\infty}^{+\infty}{\rm
d}y'\,g^*(y)M(y,y')g(y').
\end{equation}
A straightforward calculation shows that the kernel $M$ is given by
\begin{equation}
M(y,y')={1\over 8\pi^2}\,{1\over \sinh\,v}\,\ln{{\rm e}^u+{\rm e}^v\over {\rm
e}^u+{\rm e}^{-v}},
\end{equation}
where $u=(1/2)(y+y')$, $v=(1/2)(y-y')$. $M$ is positive and the following
upper bounds are easily established:
\begin{equation}
M(y,y')<\left\{\begin{array}{l}
\d {1\over4\pi^2}{\rm e}^{-u}\\[8mm]
\d {1\over4\pi^2}\,{v\over \sinh\,v}. \end{array}\right.
\end{equation}
The integral in (D.14) is split into two parts. the first one is the
restriction of the integral to values of $u$ which are larger than some $L$
($L>0$); the second part comes from $u<L$. The first and second bounds in
(D.16) are used for $u>L$ and $u<L$ respectively:
\begin{eqnarray}
\hspace*{-1.3cm}
I_{u>L}&\hspace*{-2mm}=&\hspace*{-2mm}{1\over 2\pi^2}\,{\rm
e}^{-L}\int_0^\infty{\rm d}u\,{\rm e}^{-
u}\int_{-\infty}^{+\infty}{\rm d}v\,g^*(L+u+v)g(L+u+v)\nonumber\\
&\hspace*{-2mm}<&\hspace*{-2mm} {1\over 2\pi^2}\,{\rm e}^{-L}\int_0^\infty{\rm
d}u\,{\rm e}^{-u}
\left[\int_{-\infty}^{+\infty}{\rm d}v\,\left|g(L+u+v)\right|^2
\int_{-\infty}^{+\infty}{\rm d}v\,\left|g(L+u-v)\right|^2\right]^{1\over
2}\nonumber\\
&\hspace*{-2mm}=&\hspace*{-2mm}{1\over 2\pi^2}\,{\rm e}^{-L}\Vert g\Vert^2.
\end{eqnarray}
For the second part of $I$ we write
\begin{eqnarray}
I_{u<L}&<& {1\over 2\pi^2}\int_{-\infty}^{+\infty}{\rm d}v\,{v\over \sinh\,v}
\int_0^\infty{\rm d}u\,\left|g(L-u-v)g(L-u+v)\right|\nonumber\\ &<& {1\over
\pi^2}\int_{-\infty}^{+\infty}{\rm d}v\,{v\over \sinh\,v} \left[\int_{-
\infty}^{L-v}{\rm d}y\,\left|g(y)\right|^2 \int_{-\infty}^{L+v}{\rm
d}y\,\left|g(y)\right|^2\right]^{1\over 2}\nonumber\\ &=&{1\over
4}\left[\int_{-\infty}^L{\rm d}y\left|g(y)\right|^2\right]^{1\over 2}\Vert
g\Vert. \end{eqnarray}

Similar results hold for $J$. The inequality (5.8) is obtained from (D.11) and
(D.18).

{\Large\bf Figure caption}

\begin{enumerate}
\item[Fig.~1.] The dilaton black hole space-time in the Kruskal coordinates
$(z^+,z^-)$. The infalling matter creating the black hole follows the line
$z^+=z_0^+$. The  curvature is singular on $C$; $H$ is the black hole horizon.
The observer defined in Appendix~A describes the world line $O$
of equation $z^+(z^-+\Delta)=-z_0^+\Delta\lambda Y$. Although $\lambda Y$
has to be large, the observer detects the thermal horizon radiation at late
eigentimes.
\end{enumerate}

\vspace{1cm}
\begin{thebibliography}{99}
\bibitem{Cal} C.G. Callan, S.B. Giddings, J.A. Harvey and A. Strominger,
Phys. Rev. D {\bf 45}, R 1005 (1992).
\bibitem{Int2} For recent reviews, see S.B. Giddings, in the Proceedings of
the International Workshop on Theoretical Physics, 1993, Erice, Italy (World
Scientific, 1993); J.A.~Harvey and A. Strominger, in the Proceedings of the
1992 TASI Summer School, Boulder, Colorado (World Scientific, 1993).
\bibitem{Int3} S.B. Giddings and W.M. Nelson, Phys. Rev. D {\bf 46}, 2486
(1992).
\bibitem{Int4}G. Morchio and F. Strocchi, Ann. Inst. H. Poincar\'e, {\bf 33},
251 (1980); F. Strocchi, Lecture Notes, S.I.S.S.A. Report No.~213/FM (1992).
\bibitem{Int5} S.W. Hawking, Commun. Math. Phys. {\bf 43}, 199 (1975);
R.M.~Wald, Commun. Math. Phys. {\bf 45}, 9 (1975); L.~Parker, Phys. Rev. D
{\bf 12}, 1519 (1975); S.B.~Giddings and W.M.~Nelson, Ref.~\cite{Int3}.
\bibitem{Int6} Th. Gallay and G. Wanders, Helv. Phys. Acta, {\bf 66}, 378
(1993).
\bibitem{Int7} T. Jacobson, Phys. Rev. D {\bf 48}, 728 (1993).
\bibitem{Int8} R.Haag, H. Narnhofer and U. Stein, Commun. Math. Phys. {\bf
94}, 219 (1984); B.S.~Kay and R.M.~Wald, Phys. Rep. {\bf 207}, 49 (1991).
\bibitem{Int9} K. Fredenhagen and R. Haag, Commun. Math. Phys. {\bf 108}, 91
(1987).
\end{thebibliography}
\end{document}